\documentclass{svmult}

\usepackage{makeidx}     
\usepackage{graphicx}    
\usepackage{multicol}    

\makeindex             

\usepackage{amssymb}
\usepackage{amsmath}
\usepackage{comment}
\usepackage{graphicx}
\usepackage{dcolumn}
\usepackage{bm}
\usepackage{psfig,epsfig}

\newcommand{\NPnote}[1]{}

\newcommand{\bequ}[1]{\begin{equation}\label{#1}}
\newcommand{\eequ}{\end{equation}}
\newcommand{\barr}[1]{\begin{eqnarray}\label{#1}}
\newcommand{\earr}{\end{eqnarray}}
\newcommand{\barrz}{\begin{eqnarray*}}
\newcommand{\earrz}{\end{eqnarray*}}

\newcommand{\vecta}{{\mathbf{a}}}

\newcommand{\vecte}{{\mathbf{e}}}
\newcommand{\vectf}{{\mathbf{f}}}

\newcommand{\vectk}{{\mathbf{k}}}

\newcommand{\vectp}{{\mathbf{p}}}

\newcommand{\vectu}{{\mathbf{u}}}

\newcommand{\vectx}{{\mathbf{x}}}

\newcommand{\calL}{{\mathcal{L}}}

\newcommand{\calP}{{\mathcal{P}}}

\newcommand{\bbZ}{{\mathbb{Z}}}

\newcommand{\frakp}{{\mathfrak{p}}}

\newcommand{\rmi}{{\mathrm{i}}}

\newcommand{\re}{\mathsf{Re}}

\begin{document}



\title*{Low-wavenumber forcing and turbulent energy dissipation}
\author{Charles R. Doering\and
Nikola P. Petrov}
\institute{Department of Mathematics and Michigan Center 
for Theoretical Physics \\
University of Michigan, Ann Arbor, MI 48109, USA \\
E-mail addresses:  \texttt{doering@umich.edu}
and \texttt{npetrov@umich.edu}}
%
%
\maketitle

\section{Introduction}  \label{sec:intro}

In many Direct Numerical Simulations (DNS) 
of turbulence researchers inject power into the fluid 
at large scales and then observe 
how it ``propagates'' to the small scales 
\cite{%
JimenezWSR93,%
GrossmannL94,%
Sreenivasan95,%
WangCBW96,%
YeungZ97,%
SreenivasanA97,%
Sreenivasan98,%
CaoCD99,%
YamazakiIK02,%
GotohFN02,%
KanedaI03,%
SchumacherSY03%
}.
One such type of stirring is to take the force $\vectf(\vectx,t)$ 
to be proportional to the projection 
of the velocity $\vectu(\vectx,t)$ of the flow 
onto its lowest Fourier modes, 
while keeping the rate of injected external power constant.  
In this paper we perform a simple but rigorous analysis 
to establish bounds on the relationship 
between the energy dissipation rate 
(which is the same as the injected power) 
and the resulting Reynolds number.  
While this analysis cannot give detailed information 
of the energy spectrum, it does provide some indication 
of the balance of energy between the lower, 
directly forced, modes, and those excited by the cascade.  
This work is an extension of the analysis in 
\cite{ChildressKG01,DoeringF02,DoeringES03}, 
where the force is fixed (not a functional of the velocity).  

Consider fluid in a periodic $d$-dimensional box 
of side length~$\ell$.  
The allowed wave vectors $\vectk$ are of the form 
$\vectk=\frac{2\pi}{\ell}\vecta$, 
where $\vecta\in\bbZ^d$ is a $d$-dimensional vector 
with integer components.  
Let $\calL$ be the subset of wave vectors 
that have the smallest possible wavenumber 
(namely, $\frac{2\pi}{\ell}$); 
$\calL$ consists of $2d$ elements: 
$\calL = \{\pm\frac{2\pi}{\ell}\vecte_1,\,
	\ldots,\,\pm\frac{2\pi}{\ell}\vecte_d\}$.  
The operator $\calP$ projects the vector field 
\[
\vectu(\vectx,t) 
= \sum_\vectk \hat\vectu(\vectk,t) \, 
        \E^{\rmi \vectk\cdot\vectx} 
\]
onto the subspace spanned 
by the Fourier components with wave vectors in~$\calL$: 
\begin{equation}  \label{eq:P-intro}
\calP\vectu(\vectx,t) 
= \sum_{\vectk\in\calL} \hat\vectu(\vectk,t) \, 
        \E^{\rmi \vectk\cdot\vectx} \ .  
\end{equation}
Obviously, $\calP$ maps $L^2$ into $L^2$ vector fields; 
in fact, $\calP\vectu$ is $C^\infty$ in the spatial variables.  
The projection also preserves the incompressibility property.  
That is, if $\nabla\cdot\vectu(\vectx,t)=0$, 
then $\nabla\cdot\calP\vectu(\vectx,t)=0$.  

The Navier-Stokes equation is
\begin{equation}  \label{eq:NS}
\dot\vectu + (\vectu\cdot\nabla)\vectu + \frac1\rho\nabla p 
= \nu \Delta \vectu + \vectf \ , 
\end{equation}
with $\vectf(\vectx,t)$ taken in the form 
\begin{equation}  \label{eq:f-def}
\vectf(\vectx,t) 
= \epsilon \, 
        \frac{\calP\vectu(\vectx,t)}
                {\frac1{\ell^d} \|\calP\vectu(\cdot,t)\|_2^2} \ .  
\end{equation}
where $\|\cdot\|_2$ stands for the $L^2$-norm, 
$\|\calP\vectu(\cdot,t)\|_2 := 
\left[\int |\calP\vectu(\vectx,t)|^2 \,\D^d\vectx \right]^{\frac12}$.

This choice of forcing ensures that the input power is constant:  
\begin{equation}  \label{eq:power}
\int  \vectu(\vectx,t) \cdot \vectf(\vectx,t) \, \D^d\vectx 
         = \ell^d\epsilon \ .
\end{equation}
In this approach $\epsilon$, $\nu$ and $\ell$ 
are the (only) control parameters.  
On average, the power input is the viscous energy dissipation rate: 
\begin{equation}  \label{eq:eps-def}
\epsilon 
:= 
\frac{1}{\ell^d} 
\int  \vectu(\vectx,t) \cdot \vectf(\vectx,t) \, \D^d\vectx 
= 
\nu \, \frac{\langle\|\nabla\vectu\|_2^2\rangle}{\ell^d} \ ,
\end{equation}  
where $\langle\cdot\rangle$ stands for the long time average.  
The non-dimensional measure of energy dissipation is defined as 
\begin{equation}  \label{eq:beta-def}
\beta := \frac{\epsilon\ell}{U^3} 
\ ,
\end{equation}
which is a function of $\re:=\frac{U\ell}{\nu}$, 
the Reynolds number, 
where $U$ is the r.m.s.\ velocity defined by 
$U^2 := \frac{\langle\|\vectu\|_2^2\rangle}{\ell^d}$, 
a measure of the total kinetic energy of the fluid.  
Our analysis will establish limits on the relationship 
between $\beta$ and~$\re$.  

Because we will study the ``low-$k$'' Fourier modes 
(i.e., modes with wave vectors in $\calL$), 
we also introduce the r.m.s.\ velocity $V$ 
contained in these modes, 
\begin{equation}  \label{eq:V-def} 
V^2 := \frac{\langle\|\calP\vectu\|_2^2\rangle}{\ell^d} \ .  
\end{equation}
The bounds on the dissipation $\beta$ 
will be in terms of $\re$ and the quantity 
\begin{equation}  \label{eq:p-def}
\frakp := \frac{V}{U} 
\sim \sqrt{\frac{\mbox{``low-$k$'' kinetic energy of the fluid}}
	{\mbox{Total kinetic energy of the fluid}}} \ .  
\end{equation}
The case $\frakp\approx 1$ corresponds to laminar flow, 
when the turbulent cascade is inoperative.


\section{Derivation of the bounds}  \label{sec:main}


\subsection{Lower bounds on the energy dissipation}

To obtain lower bounds on the energy dissipation, 
we proceed as usual by multiplying the Navier-Stokes equation 
\eqref{eq:NS} by $\vectu(\vectx,t)$ and integrating 
over the volume of the fluid to obtain 
the instantaneous power balance, 
\begin{equation}   \label{eq:en-bal}
\frac12 \frac{\D}{\D t} \, \|\vectu(\cdot,t)\|_2^2 
= 
- \nu \, \|\nabla\vectu(\cdot,t)\|_2^2 + \ell^d \epsilon \ , 
\end{equation}
where 
$\|\nabla\vectu(\cdot,t)\|_2^2 :=  
\int \left|\sum_{j,m=1}^{d} 
	\partial_j u_m(\vectx,t)\right|^2 \D^d\vectx$.  

Now we use the facts that 
the lengths of wavevectors $\vectk\notin\calL$ 
are at least $2\pi\sqrt{2}/\ell$, 
and that 
$\|\vectu(\cdot,t)-\calP\vectu(\cdot,t)\|_2^2 = 
	\|\vectu(\cdot,t)\|_2^2 - \|\calP\vectu(\cdot,t)\|_2^2$, 
to derive a lower bound on $\|\nabla\vectu(\cdot,t)\|_2^2$:  
\begin{eqnarray}   \label{eq:bound-grad1}
\|\nabla\vectu(\cdot,t)\|_2^2 
&=&
\int  |\nabla\vectu(\vectx,t)|^2 \, \D^d \vectx 
= 
\ell^d \sum_{\vectk} k^2 |\hat\vectu(\vectk,t)|^2 
	\nonumber \\[2mm]
&=& 
\ell^d \, 
	\left( 
	\sum_{\vectk\in\calL} k^2 |\hat\vectu(\vectk,t)|^2 
	+ \sum_{\vectk\notin\calL} k^2 |\hat\vectu(\vectk,t)|^2 
	\right) 
	\nonumber \\[2mm]
&\geq& 
\ell^d \, 
\frac{4\pi^2}{\ell^2} 
	\left( 
	\sum_{\vectk\in\calL} |\hat\vectu(\vectk,t)|^2 
	+ 2 \sum_{\vectk\notin\calL} |\hat\vectu(\vectk,t)|^2 
	\right)  \nonumber \\[2mm]
&=& 
\frac{4\pi^2}{\ell^2} 
	\left(
	\|\calP\vectu(\cdot,t)\|_2^2 
	+ 2 \, \|\vectu(\cdot,t)-\calP\vectu(\cdot,t)\|_2^2 
	\right) 
	\nonumber \\[2mm]
&=& 
\frac{4\pi^2}{\ell^2} 
	\left(
		2\, \|\vectu(\cdot,t)\|_2^2 
		- \|\calP\vectu(\cdot,t)\|_2^2 
	\right)  \ .
\end{eqnarray}
From \eqref{eq:en-bal} and \eqref{eq:bound-grad1} 
we obtain the differential inequality 
\[
\frac12 \frac{\D}{\D t} \, \|\vectu(\cdot,t)\|_2^2 
\leq 
- \nu \, \frac{4\pi^2}{\ell^2} \, \|\vectu(\cdot,t)\|_2^2 
+ \ell^d \epsilon \ ,
\]
from which, using Gronwall's inequality, we deduce 
\begin{equation}   \label{eq:bound-u2}
\frac12 \, \|\vectu(\cdot,t)\|_2^2 
\leq 
\frac12 \, \|\vectu(\cdot,0)\|_2^2 \, 
\E^{-\frac{8\pi^2\nu}{\ell^2}t} 
+ \ell^d \epsilon \frac{\ell^2}{8\pi^2\nu} 
  \left(1-\E^{-\frac{8\pi^2\nu}{\ell^2}t} \right) \ . 
\end{equation}
The inequality \eqref{eq:bound-u2} implies that 
$\|\vectu(\cdot,t)\|_2^2$ is bounded uniformly in time, 
which in turn implies that the time average of the time derivative 
in \eqref{eq:en-bal} vanishes. 
This ensures that the time-averaged power balance 
(assuming that the limit associated 
with the long time average exists) 
is indeed given by \eqref{eq:eps-def}.  

Taking the time average of \eqref{eq:bound-grad1}, 
we obtain the bound 
\[
\frac{4\pi^2\nu}{\ell^2} \, (2 U^2 - V^2 )  \leq \epsilon \ ,
\]
which in non-dimensional variables reads 
\begin{equation}  \label{eq:beta-Lbound}
\frac{4\pi^2}{\re} (2-\frakp^2) \leq \beta \ .
\end{equation}


\subsection{Upper bound on the energy dissipation}

To derive an upper bound on $\beta$, 
we multiply the Navier-Stokes equation \eqref{eq:NS} 
by $\frac{\calP\vectu(\vectx,t)}{\|\calP\vectu(\cdot,t)\|_2}$ 
and integrate.  The term with $\dot\vectu$ gives 
a total time derivative, 
\[
\int \dot\vectu\cdot\frac{\calP\vectu}{\|\calP\vectu\|_2} 
        \,\D^d \vectx 
= 
\frac{1}{\|\calP\vectu\|_2} 
        \int \frac{\partial}{\partial t}(\calP\vectu) 
                \cdot\calP\vectu \,\D^d \vectx 
= 
\frac12 \frac{\D \, \|\calP\vectu(\cdot,t)\|_2 }{\D t}  \ .  
\]
For the viscosity term we obtain (integrating by parts) 
\[
\nu \int (\Delta\vectu)\cdot \frac{\calP \vectu} {\|\calP\vectu\|_2} 
        \,\D^d\vectx 
= 
- \nu \, \frac{\|\nabla\calP\vectu\|_2^2}{\|\calP\vectu\|_2} 
= 
- \nu \, \frac{4\pi^2}{\ell^2} \, \|\calP\vectu\|_2 \ , 
\]
while the forcing term gives 
$\ell^d \epsilon /\|\calP\vectu(\cdot,t)\|_2$ 
(cf.~\eqref{eq:power}).  

To estimate the inertial term, we introduce temporarily 
the notation 
$\vectp(\vectx,t) := \calP\vectu(\vectx,t)$.  
We will make use of the uniform (in $\vectx$ and $t$) estimate 
\begin{eqnarray}    \label{eq:sup}
\!\!\!\!\!\!\!\!\!\!
\frac{|\partial_j p_m(\vectx,t)|}{\|\vectp(\cdot,t)\|_2} 
\leq 
\sum_{\vectk\in\calL} 
        \frac{|k_j|\,|\hat u_m(\vectk,t)|}{\|\vectp(\cdot,t)\|_2} 
\leq 
\frac{2\pi}{\ell^{1+\frac{d}{2}}} \,
\frac{\sum_{\vectk\in\calL}|\hat\vectu(\vectk,t)|}
{\sqrt{\sum_{\vectk'\in\calL}|\hat\vectu(\vectk',t)|^2}}
\leq 
\frac{2\pi\sqrt{d}}{\ell^{1+\frac{d}{2}}} \ .  
\end{eqnarray}
Then the inertial term may be bounded (we use $\nabla\cdot\vectp=0$):  
\begin{eqnarray}  \label{eq:bound-inert1}
\!\!\!\!\!\!\!\!\!\!
\left|
\int 
[(\vectu\cdot\nabla)\vectu] \cdot \frac{\vectp}{\|\vectp\|_2} 
	\, \D^d\vectx 
\right| 
&=& 
\left|
\int 
\vectu \cdot \frac{\nabla\vectp}{\|\vectp\|_2} \cdot \vectu 
	\, \D^d\vectx 
\right|   
\nonumber \\[2mm]
&\leq& 
\frac{2\pi\sqrt{d}}{\ell^{1+(d/2)}} 
\int |\vectu|^2 \, \D^d\vectx 
= 
\frac{2\pi\sqrt{d}}{\ell^{1+(d/2)}} 
\, \|\vectu\|_2^2 
\ .
\end{eqnarray}
This estimate, however, is obviously 
not going to be tight for small $\re$, 
when the flow is not very turbulent.  
To improve this estimate so that 
it take into account the fact that 
for small $\re$ the energy does not ``propagate'' 
much from the large to the small wavenumbers, 
we split the velocity $\vectu$ into 
a ``low-$k$'' component, $\calP\vectu$, 
and a ``high-$k$'' one, $\vectu-\calP\vectu$.  
We will still use the uniform estimate \eqref{eq:sup} 
as well as the inequality 
$ab\leq\frac12(za^2+\frac1zb^2)$ 
which holds for any $z>0$:  
\begin{eqnarray}   \label{eq:bound-inert2}
& & 
\left|
\int 
[(\vectu\cdot\nabla)\vectu] \cdot \frac{\vectp}{\|\vectp\|_2} 
	\, \D^d\vectx 
\right| 
= 
\left|
\int 
[\vectp+(\vectu-\vectp)] \cdot \frac{\nabla\vectp}{\|\vectp\|_2} 
	\cdot [\vectp+(\vectu-\vectp)] \, \D^d\vectx 
\right|   
\nonumber  \\[2mm]
& & 
\hspace*{10mm}
\leq 
\frac{2\pi\sqrt{d}}{\ell^{1+(d/2)}} 
\int 
\left(
2\,|\vectp| \, |\vectu-\vectp| + |\vectu-\vectp|^2 
\right) \, \D^d\vectx 
\nonumber  \\[2mm]
& &
\hspace*{10mm}
\leq 
\frac{2\pi\sqrt{d}}{\ell^{1+(d/2)}} 
\int 
\left[
z|\vectp|^2 + \left(\textstyle{\frac1z+1}\right) \, |\vectu-\vectp|^2 
\right] \, \D^d\vectx 
\nonumber  \\[2mm]
& &
\hspace*{10mm}
\leq 
\frac{2\pi\sqrt{d}}{\ell^{1+(d/2)}} 
\left[
\left(\textstyle{\frac1z+1}\right) \, \|\vectu\|_2^2 
	+ \left(\textstyle{z-\frac1z-1}\right) \|\vectp\|_2^2 
\right]
\ .
\end{eqnarray}

Putting together \eqref{eq:bound-inert1} 
and \eqref{eq:bound-inert2}, we find 
\begin{eqnarray}   \label{eq:bound-inert3}
\ell^d\epsilon \frac{1}{\|\calP\vectu(\cdot,t)\|_2} 
&\leq& 
\frac12 \frac{\D \, \|\calP\vectu(\cdot,t)\|_2}{\D t} 
\nonumber  \\[1mm]
&+& 
	\frac{2\pi\sqrt{d}}{\ell^{1+(d/2)}} \, 
	\min\left\{
		\|\vectu\|_2^2,\, 
		\left(\textstyle{\frac1z+1}\right) \|\vectu\|_2^2  
		+ \left(\textstyle{z-\frac1z-1}\right) \|\calP\vectu\|_2^2
		\right\} 
\nonumber \\[2mm]
&+& 
	\nu \,\frac{4\pi^2}{\ell^2} \, 
                \|\calP\vectu(\cdot,t)\|_2 
	\ .
\end{eqnarray}

Now take the time average of all terms in the above inequality.  
First note that the average 
of the time derivative of $\|\calP\vectu(\cdot,t)\|_2$ 
gives zero due to the boundedness of 
$\|\calP\vectu(\cdot,t)\|_2$ 
(which follows from the boundedness of $\|\vectu(\cdot,t)\|_2$; 
see \eqref{eq:bound-u2}).  
To estimate the other terms, 
we use Jensen's inequality: 
if a function $\theta$ is convex 
and $\langle \cdot \rangle$ stands for averaging, 
then 
$\langle \theta\circ g \rangle \geq 
        \theta\left(\langle g \rangle\right)$ 
for any real-valued function~$g$.  
Applying this inequality to the case 
$g(t) = \|\calP\vectu(\cdot,t)\|_2$ 
and the convex function $\theta(t) = t^2$, 
we obtain (same as Cauchy-Schwarz) 
\[
\left\langle \|\calP\vectu\|_2 \right\rangle \leq 
        \sqrt{\left\langle \|\calP\vectu\|_2^2 \right\rangle }
        = \ell^{d/2} \, V  \ .
\]
On the other hand, if we take $\theta(t) = \frac{1}{t}$ for $t>0$, 
we deduce 
\[
\left\langle \frac{1}{\|\calP\vectu\|_2} \right\rangle 
\geq 
\frac{1}{\left\langle \|\calP\vectu\|_2 \right\rangle} 
\geq 
\frac{1}{\sqrt{\left\langle \|\calP\vectu\|_2^2 \right\rangle}} 
= 
\frac{1}{\ell^{d/2}\, V} \ .
\]
Plugging these estimates into \eqref{eq:bound-inert3}, 
we obtain 
\[
\epsilon 
\leq 
\nu \frac{4\pi^2}{\ell^2} V^2 
+ \frac{2\pi\sqrt{d}}{\ell} \,
	\min\left\{
		U^2 V,\, 
		\left(\textstyle{\frac1z+1}\right) U^2V 
		+ \left(\textstyle{z-\frac1z-1}\right) V^3 
		\right\} 
\ .
\]
In terms of the non-dimensional energy dissipation rate 
\eqref{eq:beta-def}, 
we can rewrite this inequality in the form 
\begin{equation}   \label{eq:beta-Ubound}
\beta 
\leq 
\frac{4\pi^2}{\re} \, \frakp^2 
+ 
2\pi\sqrt{d} \, \phi(\frakp,z) \ ,
\end{equation}
where we have introduced the function 
\begin{equation}   \label{eq:phi-def}
\phi(\frakp, z) 
:= 
\min\left\{
	\frakp,\, 
	\left(\textstyle{\frac1z+1}\right) \frakp
	+
	\left(\textstyle{z-\frac1z-1}\right) \frakp^3 
\right\} 
\ .
\end{equation}


\subsection{Compatibility of the lower and upper bounds on $\beta$}

Assembling the lower and upper bounds 
\eqref{eq:beta-Lbound} and \eqref{eq:beta-Ubound}, 
we have 
\begin{equation}  \label{eq:beta-LUbound}
\frac{4\pi^2}{\re} (2-\frakp^2) 
\leq 
\beta 
\leq 
\frac{4\pi^2}{\re} \, \frakp^2 
+ 
2\pi\sqrt{d} \, 
\phi(\frakp,z) 
\ .
\end{equation}

The compatibility of the two bounds on $\beta$ 
imposes restrictions on the allowed range of $\frakp$, 
namely, $\frakp$ should satisfy the inequality 
\begin{equation}  \label{eq:cond-p}
\frakp^2 
+ \frac{\sqrt{d}\,\re}{4\pi} \, \phi(\frakp,z)
- 1 \geq 0 \ . 
\end{equation}  
In the interval $\frakp\in[0,1]$, 
this inequality is satisfied for 
$\frakp\in[p_\mathrm{min}(\re,z),1]$, 
where 
$p_\mathrm{min}(\re,z)\approx\frac{4\pi}{\sqrt{d}\,\re}$ 
for large $\re$. 
Clearly, the lower bound on the range of $\frakp$ 
is more meaningful for smaller~$\re$.  


\subsection{Optimizing the upper bound on $\beta$}

Since we do not have {\em a priori} control over $\frakp$, 
we will derive an upper bound for $\beta$ 
by maximizing the upper bound in \eqref{eq:beta-LUbound} 
over $\frakp$, after which we use the freedom 
in the choice of the parameter $z>0$ to minimize 
for any given $\re$, which results in 
\begin{equation}  \label{eq:bound-opt}
\beta 
\leq 
\min_{z>0} 
\max_{\frakp\in[p_\mathrm{min}(\re,z),1]} 
\left[
	\frac{4\pi^2}{\re} \, \frakp^2 
	+ 
	2\pi\sqrt{d} \, \phi(\frakp,z) 
\right]
\ .
\end{equation}
Although this procedure is not difficult to implement numerically, 
we will analyze only the case of high $\re$ 
where the analysis can be carried out analytically.  
First notice that for high $\re$, the lower bound 
$p_{\mathrm{min}}(\re,z)$ is very small, 
so the maximization over $\frakp$ can be taken 
in the entire interval $[0,1]$.  
Thus 
$\phi(\frakp,z) \leq \phi^*(z) := 
	\max_{\frakp\in[0,1]} \phi(\frakp,z) 
	= (1+z-z^2)^{-1/2}$ 
for $z\in[0,\frac{1+\sqrt{5}}{2})$.  
Since for large $\re$ 
the $\re$-independent term in the right-hand side 
of \eqref{eq:bound-opt} is dominating, 
we have the high-$\re$ estimate 
\begin{equation}   \label{eq:bound-approx}
\beta
\leq 
\min_{z\in[0,\frac{1+\sqrt{5}}{2})} 
\left[
	\frac{4\pi^2}{\re} \, \phi^*(z) ^2
	+ 
	2\pi\sqrt{d} \, \phi^*(z) 
\right]
= 
\frac{16\pi^2}{5\re} + \frac{4\pi\sqrt{d}}{\sqrt{5}} 
\ .
\end{equation}
At high $\re$, the value or $\frakp$ 
maximizing $\phi(\frakp,z)$ is $\frac{2}{\sqrt{5}}$.  
We remark that it is not difficult to prove 
that the upper bound \eqref{eq:bound-approx} 
is optimal (i.e., coincides with \eqref{eq:bound-opt}) 
for $\re \geq \frac{8\pi}{3\sqrt{5d}}$.  




\section{Discussion} 

\begin{figure}
\centering
\includegraphics[width=0.5\textwidth,angle=-90]{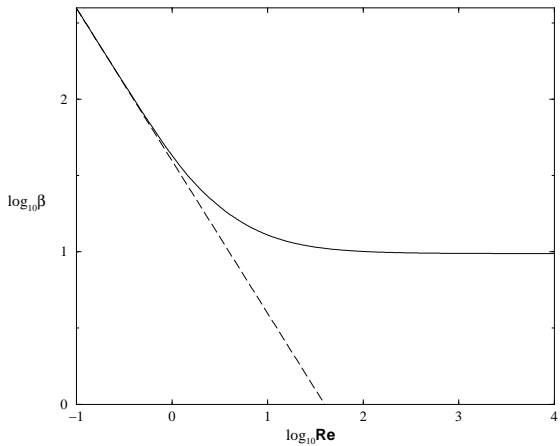}
\caption{Upper and lower bounds on $\beta$ 
(solid and dashed lines, resp.)}
\label{fig:comparison}
\end{figure}

In dimension 3, the scaling of the upper bound is in accord 
with conventional turbulence theory: 
at high $\re$, $\epsilon\sim \frac{U^3}{\ell}$ 
is independent of the molecular viscosity.  
For the type of forcing considered here, we find 
$\beta \leq 4\pi\sqrt{\frac{3}{5}} \approx  9.7339\ldots$.  
A plot of the bounds is presented in Figure~\ref{fig:comparison}.  
At low $\re$, the upper and lower bounds 
converge to each other.  
While it is difficult to compare these bounds 
quantitatively with DNS results, we note from 
\cite{Sreenivasan98} that at high $\re$, 
values of $\beta$ are typically around~$1$.  
Hence, our rigorous analysis, while yielding 
the expected scaling, overestimates 
the constants by about an order of magnitude.  

In the 3-dimensional case, 
if we assume that the cascade is Kolmogorov, 
i.e., the spectral density of the energy is 
given by $E_{\mathrm{K}}(k) = C \epsilon^{2/3} k^{-5/3}$, 
we can estimate the ``Kolmogorov'' value $\frakp_{\mathrm{K}}$ 
as follows: 
\[
E_{\mathrm{kin,\ total}} 
\approx 
 \int_{2\pi/\ell}^{\infty} E_{\mathrm{K}}(k) \, \D k \ , 
\quad 
E_{\mathrm{kin,\ low\,}k}
\approx 
\frac{2\pi}{\ell} \, E_{\mathrm{K}}\left(\frac{2\pi}{\ell}\right) 
\ ,
\]
which gives $\frakp_{\mathrm{K}} \approx \sqrt{\frac{2}{3}}$.  
Plugging this value in \eqref{eq:bound-opt} and minimizing over $z$, 
we obtain the (approximate) estimate 
\[
\beta 
\leq 
\frac{8\pi^2}{3\,\re} + 2\sqrt{2}\pi 
\approx 
\frac{26.3}{\re} + 8.9 \ ,
\]
which gives a slight improvement compared with 
the bounds \eqref{eq:bound-approx}.


\section*{Acknowledgments}
 
CRD thanks B.\ Eckhardt, J.\ Schumacher, D.\ Lohse, K.\ Sreenivasan 
for stimulating and helpful discussions.  
This work was supported in part by NSF Award PHY-0244859.





\bibliographystyle{unsrt}  
\bibliography{Doering}


\printindex
\end{document}